\documentclass[usegraphicx,usenatbib,twoside]{mn2e}
\usepackage{times}
\renewcommand\[{\begin{equation}}
\renewcommand\]{\end{equation}}

\catcode`\@=11
\def\gsim{\ifmmode{\mathrel{\mathpalette\@versim>}}
    \else{$\mathrel{\mathpalette\@versim>}$}\fi}
\def\lsim{\ifmmode{\mathrel{\mathpalette\@versim<}}
    \else{$\mathrel{\mathpalette\@versim<}$}\fi}
\def\@versim#1#2{\lower 2.9truept \vbox{\baselineskip 0pt \lineskip
    0.5truept \ialign{$\m@th#1\hfil##\hfil$\crcr#2\crcr\sim\crcr}}}
\catcode`\@=12
\def\rhos{\rho_*}
\def\Ms{M_*}
\def\As{A_*}
\def\rs{r_*}
\def\rhot{\rho_{\rm T}}
\def\rhoh{\rho_{\rm h}}
\def\MR{{\cal R}}
\def\qm{q_{\rm m}}
\def\Qm{Q_{\rm m}}
\def\srad{\sigma_{\rm r}}
\def\stan{\sigma_{\rm t}}
\def\Krad{K_{\rm r*}}
\def\Ktan{K_{\rm t*}}
\def\ra{r_{\rm a}}
\def\sa{s_{\rm a}}
\def\Mt{M_{\rm T}}
\def\vc{v_{\rm c}}
\def\Sigs{\Sigma_*}
\def\Mh{M_{\rm h}}
\def\phit{\phi_{\rm T}}
\def\xv{{\bf x}}
\def\sigv{\sigma_{\rm V}}
\def\sigp{\sigma_{\rm p}}
\def\Lin{{\rm Li}}
\def\rh{r_{\rm h}}

\def\sam{\sa^{-}}
\def\sap{\sa^{+}}
\def\cond{{\cal C}}
\def\condm{\cond_{-}}
\def\condp{\cond_{+}}
\def\reff{R_{\rm e}}

\def\vpar{v_{||}}
\def\csic{\xi_{\rm c}}

   \title[Two--component galaxy models]{Two-component galaxies with flat 
rotation curve}

   \author[Ciotti, Morganti, \& de Zeeuw]
          {Luca Ciotti $^1$, Lucia Morganti $^1$\thanks{Current address: Max-Planck-Institut f\"ur Ex. Physik, Giessenbachstra\ss{}e, D-85741 Garching, Germany}, \&  P.~Tim de Zeeuw $^{2,3}$
\\$^1$Astronomy Department, 
      University of Bologna, via Ranzani 1, 40127 Bologna, Italy
\\$^2$Leiden Observatory, Leiden University, 
      PO Box 9513, 2300 RA Leiden, The Netherlands
\\$^3$ESO, Karl Schwarzschildstrasse 2, 85748 Garching bei Muenchen, Germany}

\date{Submitted, August 9,  2008 - Resubmitted, September 25, 2008}
\pubyear{2008}

\begin{document} 
\maketitle

\begin{abstract} 

  Dynamical properties of two-component galaxy models whose stellar
  density distribution is described by a $\gamma$-model while the
  total density distribution has a pure $r^{-2}$ profile, are
  presented. The orbital structure of the stellar component is
  described by Osipkov--Merritt anisotropy, while the dark matter halo
  is isotropic.  After a description of minimum halo models, the
  positivity of the phase-space density (the model consistency) is
  investigated, and necessary and sufficient conditions for
  consistency are obtained analytically as a function of the stellar
  inner density slope $\gamma$ and anisotropy radius. The explicit
  phase-space distribution function is recovered for integer values of
  $\gamma$, and it is shown that while models with $\gamma>4/17$ are
  consistent when the anisotropy radius is larger than a critical
  value (dependent on $\gamma$), the $\gamma=0$ models are unphysical
  even in the fully isotropic case.  The Jeans equations for the
  stellar component are then solved analytically; in addition, the
  projected velocity dispersion at the center and at large radii are
  also obtained analytically for generic values of the anisotropy
  radius, and it is found that they are given by remarkably simple
  expressions.  The presented models, even though highly idealized,
  can be useful as starting point for more advanced modeling of the
  mass distribution of elliptical galaxies in studies combining
  stellar dynamics and gravitational lensing.

\end{abstract}


\section{Introduction}

Analysis of stellar kinematics (e.g. Bertin et al. 1994, Rix et
al. 1997, Gerhard et al. 2001), as well as several studies combining
stellar dynamics and gravitational lensing strongly support the idea
that the dark and the stellar matter in elliptical galaxies are
distributed so that their {\it total} mass profile is described by a
density distribution proportional to $r^{-2}$ (e.g., see Treu \&
Koopmans 2002, 2004; Rusin et al. 2003; Rusin \& Kochanek 2005;
Koopmans et al. 2006; Czoske et al. 2008; Dye et al. 2008). In
particular, Gavazzi et al. (2007), with a gravitational lensing
analysis of 22 early-type strong lens galaxies, reported a total
$r^{-2}$ density profile in the range 1-100 effective radii.  It is
clear that in this field the availability of simple dynamical models
of two-component galaxies can be useful as starting point of more
sophisticated investigations based on axysimmetric or triaxial galaxy
models (e.g., Cappellari et al. 2007, van den Bosch et al. 2008).  A
few simple yet interesting models with flat rotation curve have been
in fact constructed, such as those in which the stellar mass was
described by a power-law in a total $r^{-2}$ mass distribution
(e.g. Kochaneck 1994), or those obtained from physical arguments (in
case of disk galaxies, see e.g. Naab \& Ostriker 2007).

Here the family of two-component galaxy models whose {\it total} mass
density is proportional to $r^{-2}$, while the visible (stellar) mass
is described by the well-known $\gamma$ models (Dehnen 1993, Tremaine
et al. 1994), is presented.  Some preliminary numerical investigation
of these models has been done in Keeton (2001), and they have been
used in Nipoti et al. (2008) as diagnostics of the total mass
distribution in elliptical galaxies.  In this paper a more systematic
study of the dynamical properties of these models is presented.  It is
shown that the Jeans equations for the stellar component with
Osipkov-Merritt (Osipkov 1979, Merritt 1985, hereafter OM) radial
anisotropy can be solved analytically.  Remarkably, the projected
velocity dispersion at the center and at large radii can be expressed
in terms of the model circular velocity by means of extremely simple
formulae for generic values of the anisotropy radius and of the
central stellar density slope $\gamma$.  In principle this feature
opens the possibility to obtain preliminary indications about the
anisotropy from observations at small and large radii.  The positivity
of the phase-space density (the so-called consistency) is
investigated, by obtaining analytically the necessary and sufficient
conditions for model consistency in terms of $\gamma$, of the
anisotropy radius, and of the dark-to-stellar mass ratio within some
prescribed radius.  It is found that the phase-space distribution
function (hereafter DF) can be recovered analytically for
$\gamma=0,1,$ and $2$.  In particular, it is shown that $\gamma =0$
models in a total $r^{-2}$ density profile are unphysical for any
value of the anisotropy radius.  These results extend the class of
two-component galaxy models with explicit DF and
add to the large amount of phase-space information already available
about one and two-component $\gamma$ models (e.g., see Dehnen 1993,
Tremaine et al. 1994, Hiotelis 1994, Carollo et al. 1995, Ciotti 1996,
1999; Baes et al. 2005, Buyle et al. 2007, Ciotti \& Morganti 2008).

The paper is organized as follows. In Section 2 the main structural
properties of the models are presented, while in Section 3 an
investigation of the phase-space properties of the models is carried
out both from the point of view of necessary and sufficient conditions
for consistency and of direct recovery of the DF in specific cases. In
Section 4 the solution of the Jeans equation with OM radial anisotropy
is presented, together with their projection at small and large radii.
Finally a short summary of possible use of the present models in
observational works is given.

\section{The models}

\subsection{Stellar distribution}

The density profile of spherical $\gamma$ models is 
\[
\rhos(r)={\As\over s^{\gamma}(1+s)^{4-\gamma}},\quad 
         \As\equiv {(3-\gamma)\Ms\over 4\pi\rs^3},
\label{eq:rhos}
\]
where $0\leq\gamma <3$, $\Ms$ is the total stellar mass, $\rs$ is a
scale-length, and $s\equiv r/\rs$ is the dimensionless radius.  These
models have been investigated extensively, and here only the properties 
of present use are listed.  In particular, the
cumulative stellar mass within $r$ is given by
\[
\Ms(r)=\Ms\times\left({s\over 1+s}\right)^{3-\gamma},
\label{eq:mass}
\]
so that the dimensionless 
half-mass spatial radius is $s_{\rm h}=1/(2^{1\over 3-\gamma}-1)$.
The projected stellar surface density
\[
\Sigs(R)=2\int_R^{\infty}{\rhos(r)rdr\over\sqrt{r^2-R^2}},
\label{eq:surf}
\]
(e.g., Binney \& Tremaine 2008) 
cannot be expressed in terms of elementary functions
for generic values of $\gamma$, however the asymptotic behaviour of $\Sigs(R)$
is easily obtained for $R\to 0$
\[
\Sigs(R)\sim\As\rs\cases{
         \displaystyle{4\over (3-\gamma)(2-\gamma)(1-\gamma)},
                                                     \quad (0\leq\gamma <1);
         \cr\cr
         \displaystyle{-2\log \eta},
                       \quad\quad\quad\quad\quad\quad\quad (\gamma =1);
         \cr\cr
         \displaystyle{\sqrt{\pi}\Gamma(\gamma/2-1/2)\over\Gamma(\gamma/2)}
                       \eta^{1-\gamma},
                                                     \quad(1<\gamma\leq 3);}
\label{eq:surfz}
\]
and for $R\to\infty$ 
\[
\Sigs(R)\sim{\pi\As\rs\over2\eta^3},\quad(0\leq\gamma\leq 3).
\]
In the equations above $R$ is the radius on the projection plane,
$\eta\equiv R/\rs$ is its normalized value, and $\Gamma$ is the
complete Euler Gamma function; note that the first of
eqs.~(\ref{eq:surfz}) is the exact value of the projection integral
for $R=0$ when $0\leq\gamma <1$.

\subsection{Total and dark matter distribution}

\begin{figure}
\includegraphics[height=0.6\textheight,width=0.82\textwidth]{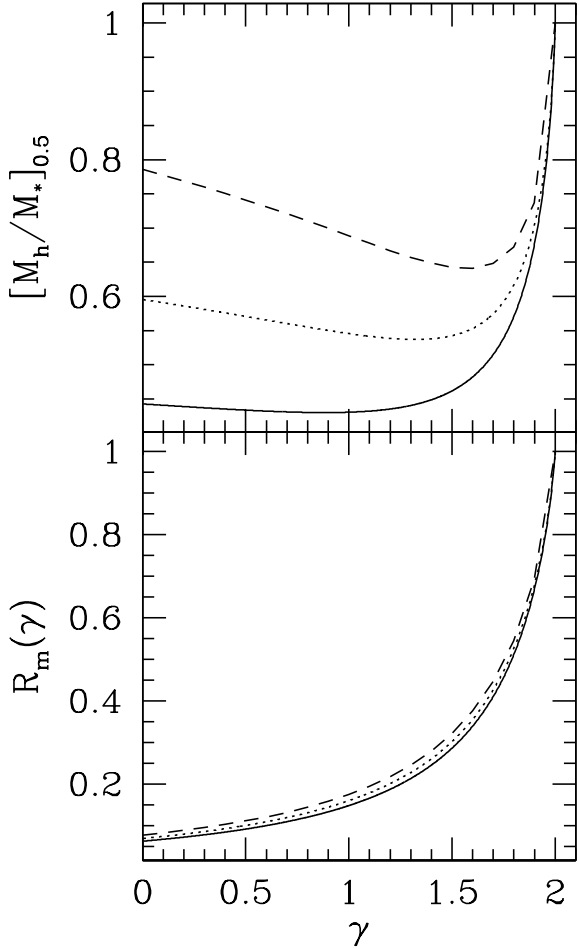}
\caption{ Bottom panel: minimum value of $\MR$ as a function of
  $\gamma$ for the halo density positivity (solid line,
  eq.~[\ref{eq:mrp}]), monotonicity (dotted line,
  eq.~[\ref{eq:mona}]), and to satisfy the WSC 
  in the isotropic case (dashed line, eqs.~[18]-[\ref{eq:wscha}]).  
  Top panel: the dark to stellar mass ratio 
  within the half-mass radius of the stellar component, 
  as given in eq.~(\ref{eq:mrm}), 
  for the three limits in the botton panel.}
\label{fig:halo}
\end{figure}

By assumption the total mass density is taken to be
\[
\rhot(r)={\MR\As\over s^2},
\label{eq:rhot}
\]
where $\MR$ is a dimensionless scale factor which measures the
importance of the dark matter density with respect to the stellar one:
therefore, the stellar distribution would be a tracer in the total
density distribution in the formal limit $\As\to0$ and $\MR\to\infty$,
in a way such that the product $\MR\As$ remains constant.  The
cumulative total mass within $r$ is
\[
\Mt(r)=4\pi\MR\As\rs^3\,s,
\label{eq:mtot}
\] 
and the system (constant) circular velocity is $\vc^2=4\pi
G\MR\As\rs^2$; from this expression the dimensionless constant $\MR$
(or the density scale $\As$) everywhere it appears in favor of $\vc$.
The total projected mass density at $R$ is obtained from
eqs.~(\ref{eq:surf}) and (\ref{eq:rhot}) as
\[
\Sigma_{\rm T}(R)={\pi\MR\As\rs\over\eta},
\label{eq:surft}
\]
so that the total mass contained within the cylinder of radius $R$ 
is
\[
M_{\rm PT}(R)=2\pi\int_0^R \Sigma_{\rm T}(R)\, R\, dR=2\pi^2\MR\As\rs^3\eta.
\label{eq:mpt}
\]
Not all values of the coefficient $\MR$ and of the inner stellar
density slope $\gamma$ are compatible. In fact a first limitation is
given by the request of positivity for the {\it halo} density
\[
\rhoh(r)={\As\over s^2}\left[
                       \MR-{s^{2-\gamma}\over (1+s)^{4-\gamma}}\right].
\label{eq:rhoh}
\]
This request restricts the value of $\gamma$ to the interval
$0\leq\gamma\leq 2$, independently of the value of $\MR$. With 
$\gamma$ in the acceptable range, $\rhoh$ is positive provided that
\[
\MR\geq\MR_m(\gamma)={4(2-\gamma)^{2-\gamma}\over (4-\gamma)^{4-\gamma}},
\label{eq:mrp}
\]
(see Appendix A). For example, $\MR_m(0)=1/16$, $\MR_m(1/2)=0.0916$,
$\MR_m(1)=4/27$, and $\MR_m(2)=1$; in Fig.~\ref{fig:halo} (bottom
panel) the minimum value $\MR_m(\gamma)$ for halo positivity is
represented by the solid line.  A dark halo with $\MR_m$ is called a
{\it minimum halo}. While the density distribution of the minimum halo
increases at the center as $r^{-2}$ for $0\leq\gamma <2$, for
$\gamma=2$ it results $\rhoh\propto r^{-1}$, and so minimum halo
$\gamma=2$ models are more and more baryon dominated near the center.
{\bf We remark that the local mass-to-light ratio, proportional to
  $\rhot(r)/\rhos(r)$ under the hypothesis of a constant stellar
  mass-to-light ratio, is a non-monotonic function of $r$ as it
  increases near the center and for $r\to\infty$. The only exception
  is represented by the $\gamma=2$ case, which is characterized by a
  monotonically increasing mass-to-light ratio for increasing $r$.}

Of course, the positivity of $\rhoh$ is just a first condition for the
acceptability of the model.  A plausible second request is the
monotonicity of $\rhoh$ as a function of radius: while at this stage
monotonicity reduces to the determination of a minimum value of
$\MR_m(\gamma)$ so that $d\rhoh/dr \leq 0$, in Section 3 it will be
shown that this request is based on deeper physical arguments than
simple structural plausibility. The explicit calculation of this
additional restriction of $\MR$ is given in Appendix A, and the
resulting function $\MR_m(\gamma)$ is shown in Fig.~\ref{fig:halo}
(bottom panel) with the dotted line: it is apparent that the request
of monotonicity is just a little bit more stringent than positivity,
and that in the $\gamma=2$ case the two requests coincide.

It can be of interest in applications to evaluate the relative 
amount of dark to visible mass within a prescribed radius.  This 
quantity is easily calculated from eqs.~(\ref{eq:mass}) and~(\ref{eq:mtot}).
For example, within the
half-mass radius $\rh$ one has
\[
{\Mh(\rh)\over\Ms(\rh)}\geq {2(3-\gamma)\MR_m(\gamma)\over
                             2^{1/(3-\gamma)}-1}-1,
\label{eq:mrm}
\]
where $\Mh(r)=\Mt(r)-\Ms(r)$. In Fig.~\ref{fig:halo} (top panel) the
mass ratios corresponding to the three limits in the bottom panel are
shown. The values are smaller than unity for all values of
$\gamma$, except for the $\gamma=2$ case. 

Another observationally relevant quantity is the projected mass ratio
of dark-to-visible, for example within the effective radius $\reff$ of
the stellar distribution. This is given by
\[
{2 M_{\rm Ph}(\reff)\over\Ms}\geq\pi(3-\gamma)\MR_m(\gamma)\eta_{\rm{e}} -1,
\quad \eta_{\rm{e}}\equiv\reff/\rs, 
\label{eq:mpe}
\]
where for example 
$\eta_{\rm{e}}\simeq 2.9036$ for $\gamma=0$ (Dehnen 1993), 
$\eta_{\rm{e}}\simeq 2.3585$ for $\gamma=1/2$, $\eta_{\rm{e}}\simeq
1.8153$ for $\gamma=1$ (Hernquist 1990), and $\eta_{\rm{e}}\simeq 0.7447$ for
$\gamma=2$ (Carollo et al. 1995; note that in Jaffe 1983 the slightly
erroneous coefficient 0.763 is reported). These values
translate into mass ratios of $\simeq 0.71$, 0.70, 0.69, and 1.34  when 
considering the minimum value of $\MR_m(\gamma)$ for halo positivity, 
for $\gamma=0,1/2,1,$ and $2$, respectively.

\section{The phase-space  distribution function}

Before solving the Jeans equations, it is useful to discuss some basic
property of the phase-space DF of the presented models, in order to
exclude physically inconsistent combinations of parameters (i.e.,
choices that would correspond to a somewhere negative
DF). Fortunately, as discussed extensively in Ciotti \& Pellegrini
(1992, hereafter CP92), C96, and C99, it is possible to obtain lower
bounds for the OM anisotropy radius as a function of the density slope
and the total mass profile, without actually recovering the DF, which
is in general impossible in terms of elementary functions.  More
specifically, in CP92 a simple theorem was proved regarding the
necessary and sufficient limitations on $\ra$ in multi-component OM
models, while more recently An \& Evans (2006) proved the so-called
``cusp slope-central anisotropy'' theorem (see also eq.~[28] in de
Bruijne et al. 1996): the link between the two results is briefly
addressed in Ciotti \& Morganti (2008).  Before using the CP92 test
some preliminary work is however in order, because at variance with
the common case of finite total mass, the total potential
\[
\phit =\vc^2\ln s
\]
is now quite peculiar, being logarithmic. This means that in principle 
orbits of any energy can be present, and the standard OM
prescription must be reformulated to take into account the divergent
behaviour of the potential both at $r=0$ and $r=\infty$.
Thus, a DF with the functional dependence 
\[
f=f(Q),\quad Q\equiv E + {J^2\over 2\ra^2},
\]
is assumed, where $E =\phit+v^2/2$ and $J$ are 
the energy and angular momentum modulus of each star (per unit mass), 
respectively.  
Note that,
at variance with the usual OM parameterization, no cut on $f$ for
negative $Q$ is present.  By integration over the velocity space it is
easy to show that for a given density component (stars or halo) in the
total potential $\phit$, the density is related to its DF by
\[
\rho={4\pi\over 1+r^2/\ra^2}\int_{\phit}^{\infty}f(Q)\sqrt{2(Q-\phit)}\;dQ;
\label{eq:OM}
\]
in principle, $\ra$ can be different for stars and dark matter.  The
analogy with the standard OM relation is apparent (e.g., see Binney \&
Tremaine 2008).  Following a similar treatment, it can also be shown 
that the radial ($\srad$) and tangential ($\stan$) components of the
velocity dispersion tensor are related as in the standard
OM case, i.e.
\[
\beta(r)\equiv{1-{\stan^2(r)\over{2\srad^2(r)}}}={r^2\over{r^2+\ra^2}},
\label{betaOM}
\]
so that the fully isotropic case is obtained for $\ra\to\infty$, while
for $\ra=0$ the galaxy is supported by pure radial anisotropy.  For
finite values of $\ra$, the velocity dispersion tensor becomes
isotropic for $r\to 0$ (in practice for $r<\ra$), and fully
radially anisotropic for $r\to\infty$ (in practice for $r>\ra$).
Introducing the augmented density
\[
\varrho(r)\equiv \rho(r)\left(1+{r^2\over\ra^2}\right),
\]
eq.~(\ref{eq:OM}) can be Abel
inverted, obtaining
\begin{eqnarray}
f(Q)&=&{1\over\sqrt{8}\pi^2}{d\over dQ}
     \int_Q^{\infty}{d\varrho\over d\phit}
     {d\phit\over\sqrt{\phit-Q}}\cr
     &=&{1\over\sqrt{8}\pi^2}
     \int_Q^{\infty}{d^2\varrho\over d\phit^2}
     {d\phit\over\sqrt{\phit-Q}},
\label{eq:OMinv}
\end{eqnarray}
where it is intended that $\varrho$ is expressed in terms of $\phit$,
and the second identity follows from integration by parts when
considering the untruncated nature of the studied density
distributions.

Moreover, for the present class of models it can be also proved that
the velocity profile (VP, e.g. Carollo et al. 1995) can be written as
\begin{equation}
\label{eq:vp}
\Sigs\mbox{VP}=4\pi\int^\infty_R {g(r,R)r dr\over\sqrt{r^2-R^2}}
\int^\infty_{\Qm}f(Q)dQ,
\end{equation}
with 
\begin{equation}
g(r,R)={\ra^2\over\sqrt{r^2+\ra^2}\sqrt{r^2+\ra^2-R^2}},
\end{equation}
and
\begin{equation}
\Qm=\phit+{r^2+\ra^2\over r^2+\ra^2-R^2}{\vpar^2\over 2},
\end{equation}
where $\vpar$ is the velocity along the line of sight direction.  
The inner integral in eq.~(\ref{eq:vp}) can be simplified 
by using the first identity in eq.~(\ref{eq:OMinv}).

\subsection{Necessary and sufficient conditions for consistency}

Repeating the same treatment of CP92, after
differentiation of eq.~(\ref{eq:OM}) with respect to $\phit$, it
follows that a {\it necessary condition} for the positivity of the DF 
is that
\[
{d\varrho (r)\over dr}\leq 0\quad{\rm[NC]}.
\label{eq:NC}
\]
This necessary condition for the DF positivity is 
independent of the radial dependence of the other density components
of the system.
\begin{figure}
\includegraphics[height=0.4\textheight,width=0.5\textwidth]{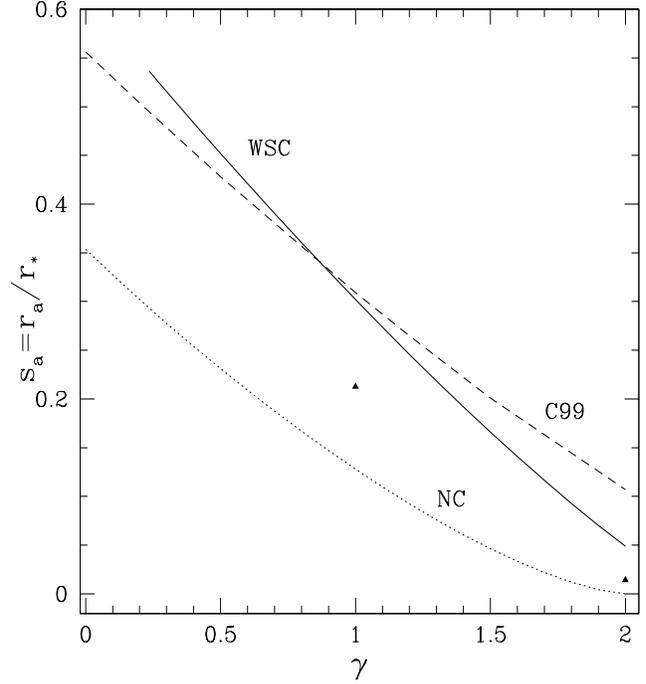}
\caption{The NC limit for consistency of $\gamma$ models
  is shown with the dotted line: all values of the anisotropy radius
  below the line correspond to inconsistent models (C99).  Solid line:
  the WSC for the present models, i.e., the locus
  above which the models are certainly consistent. The line begins at
  $\gamma=4/17$, as the WSC is not satisfied for centrally flatter
  models. Dashed line: the WSC for one-component $\gamma$ models as
  determined in C99.  The triangles are the true anisotropy limits for
  the $\gamma=1$ and $\gamma=2$ models with halo obtained 
  from their DF (Section 3.2).}
\label{fig:wsc}
\end{figure}

The CP92 {\it weak sufficient condition} for consistency is recovered
by requiring that the second derivative inside the integral in
eq.~(\ref{eq:OMinv}) be positive. In analogy with
eq.~(\ref{eq:NC}), this condition can be expressed as a 
function of radius as
\[
{d\over dr}\left[{d\varrho(r)\over dr}{r^2\over\Mt(r)}\right]\geq 0
\quad{\rm [WSC]},
\label{eq:WSC}
\]
where the total mass is given by eq.~(\ref{eq:mtot}).

\subsubsection{Isotropic halo consistency}

The first application of eqs.~(\ref{eq:NC}) and (\ref{eq:WSC})
concerns the consistency of the halo density distribution $\rhoh$.
For simplicity we restrict to the isotropic case, and then
eq.~(\ref{eq:NC}) shows the equivalence of the request of monotonicity
of $\rhoh$ discussed in Sect. 2.2 with the necessary condition 
for a phase--space consistent halo.  The WSC for a fully isotropic
halo can be discussed analytically as described in Appendix A, and the
resulting limit is represented by the dashed line in Fig.1 (bottom
panel): note how the three conditions of halo positivity, monotonicity,
and consistency produce very similar curves, that coincide for $\gamma=2$.

{\bf Of course, the restriction of the study to an isotropic halo is
  quite arbitrary, as the virialized end-states of $N$-body collapses
  are characterized by some amount of orbital anisotropy (e.g., van
  Albada 1982; Nipoti, Londrillo \& Ciotti 2006). However, the present 
  investigation is mainly  focused on the properties of the visible component, 
  and therefore we adopt the simplest dynamical structure for the halo. }

\subsubsection{Anisotropic stellar consistency}

In general, when dealing with OM anisotropic systems, the investigation of 
the NC and WSC (eqs.~[\ref{eq:NC}]-[\ref{eq:WSC}]),
and the study of the DF positivity (eq.~[\ref{eq:OMinv}]) lead to
inequalities of the kind 
\[
F+{G\over\sa^2}\geq 0,
\label{eq:cond}
\]
that must hold over all the domain of interest, which we indicate with 
$\cond$.  In practice, the functions $F$ and $G$ are 
radial functions (in the case of the NC and
WSC) or functions of the phase-space variable $Q$ (in the case of the DF),
depending on the specific context.
Following C99, here
we recall that according to eq.~(\ref{eq:cond}) {\it all} OM models 
can be divided in two main families. 
In the first case, the function $F$ is nowhere negative over
$\cond$ (this could be the case of a system with a positive isotropic
DF). Then, the {\it lower bound} to $\sa$ is given by
\[
\sam=\sqrt{{\rm max}\left[0,{\rm sup}_{\cond}\left(-{G\over F}\right)\right]},
\label{eq:sam}
\]
and the condition~(\ref{eq:cond}) is satisfied provided $\sa\geq\sam$:
in particular, if $G$ is also positive over $\cond$
then the system can be supported by radial orbits only.
However, it may happen that $F$ is positive only over some proper
subset $\condp$ of $\cond$, and negative (or zero) over $\condm$. It
trivially follows that in this second class of models
if $G$ is negative over some subset\footnote{ 
In C99 and Ciotti (2000) it is erroneously stated 
that the model is inconsistent if $G<0$ everywhere on $\condm$.
All the results presented therein are however correct.} of
$\condm$, then the condition~(\ref{eq:cond}) cannot be satisfied for any
value of $\sa$.
On the contrary, it may happen that $G$
is everywhere positive on $\condm$: in this case one must consider not
only the lower limit $\sam$ over $\condp$, 
but also the {\it upper bound}
\[
\sap=\sqrt{{\rm inf}_{\condm}\left(-{G\over F}\right)},
\label{eq:sap}
\]
so that $\sam<\sap$ over $\condm$ for consistency.  Summarizing, if
$F\geq 0$ over all its domain (i.e., $\condp=\cond$), then
$\sa\geq\sam$ satisfies inequality~(\ref{eq:cond}). If $F\leq 0$ over
some set $\condm$ but $G\geq 0$ there, then the inequality
$\sam\leq\sa\leq\sap$ must be verified. Finally, if over $\condm$
$\sap <\sam$ or $G <0$ somewhere, then inequality~(\ref{eq:cond})
cannot be satisfied and, in case of a DF analysis, the model must be
rejected as inconsistent.

For example, in the case of $\gamma$ models the $\sam(\gamma)$ limit
from the NC has been calculated analytically in C99, and the critical
value of the anisotropy radius expressed in units of $\rs$ are
$\simeq0.354$, $\simeq 0.128$, and $0$ for $\gamma=0,1,2$,
respectively.  In other words, smaller values of $\sa$ correspond to
physically inconsistent models (even though the solution of the
associated Jeans equations is positive - see the following
Section). The anisotropy limit over the whole range of $\gamma$ here
considered is represented with the dotted line in Fig.~2: note how
centrally flatter models are associated with larger values of the
critical anisotropy radius.

We now discuss the case of the WSC for the stellar component of our
models. As shown in Appendix A, simple algebra reveals that the
function $F$ in equation~(\ref{eq:cond}) is positive everywhere for
$\gamma >4/17$, and from eq.~(\ref{eq:condm}) the maximum of the
function $-G/F$ can be determined by solving an equation of degree
four.  The resulting value of $\sam(\gamma)$ is shown with the solid
line in Fig.~\ref{fig:wsc}. For $0\leq\gamma\leq 4/17$ the function
$F$ has two positive roots, delimiting the interval $\condm$ on which
$F<0$. The function $G$ is positive on $\condm$, so that we can
determine the two values $\sap$ (on $\condm$) and $\sam$ (on
$\condp$): however it turns out that $\sap <\sam$ for $0\leq\gamma\leq
4/17$, so that the WSC is not satisfied, and for this reason the solid
line interrupts in Fig.~\ref{fig:wsc}.  Of course, being just a
sufficient condition, this result does not exclude that consistent
models exist for $\gamma <4/17$, but this is not assured as it is for the
models with $\gamma >4/17$.  For reference, in Fig.~\ref{fig:wsc} the
dashed line represents the WSC for one-component $\gamma$ models as
derived in C99.  In particular, note how for models with $\gamma\gsim
1$, the presence of the halo appears to increase the model ability to
sustain radial orbital anisotropy, while flatter models in presence of
the halo are less able to sustain anisotropy.

\subsection{Explicit Phase-Space DF}

For generic $\gamma$, 
eq.~(\ref{eq:OMinv}) can be rewritten as 
\begin{eqnarray}
f(q)&=&{\As\over\sqrt{8}\pi^2\vc^3}{d\over dq}
     \int_q^{\infty}{d\tilde\varrho\over d\Psi}{d\Psi\over\sqrt{\Psi -q}}=\cr
    &&{\As\over\sqrt{8}\pi^2\vc^3}
       \left[U(q)+{V(q)\over\sa^2}\right],
\end{eqnarray}
where $\Psi\equiv\phit/\vc^2$ and $q\equiv Q/\vc^2$.  The function
$\tilde\varrho$ is the augmented density in eq.~(18) normalized to
$\As$, expressed in terms of the total potential.
This is accomplished by elimination of the radius
from the dimensionless identity $s=\exp(\Psi)$ obtained from eq.~(14). 
Not surprisingly, for generic $\gamma$ the functions $U$ and $V$ cannot be
expressed in terms of known functions, however in Appendix B it is
shown that for $\gamma=0,1$ and $2$ the functions $U$ and $V$ can be
expressed as simple linear combinations of exponentials and
Polylogarithms.  Numerical inspection of the DFs shows that for
$\gamma=1$ and 2 the isotropic component $U$ is positive for all
values of $q$, and the lower bound on the anisotropy
limit are $\sam=0.212675$ (for $\gamma=1$) and $\sam=0.0141$
(for $\gamma=2$). These two limits are represented as solid triangles
in Fig.~\ref{fig:wsc}. As expected, their position is found between
the NC and the WSC loci. Instead, in the $\gamma=0$ case the function
$U$ is negative and the function $V$ is positive 
for $q\,\lsim -2.4$: however $\sam >\sap$, and the model is inconsistent.
Note that the OM anisotropy limit for $\gamma$ models without dark
halo, derived in C99 from their DF, is $\sam(0)\simeq 0.445$, 
$\sam(1)\simeq 0.202$, and $\sa(2)=0$, so we conclude that 
{\it the presence of the DM halo slightly reduces 
the ability of the stellar density distribution to
sustain radial anisotropy}.
\begin{figure}
\includegraphics[height=0.5\textheight,width=0.55\textwidth]{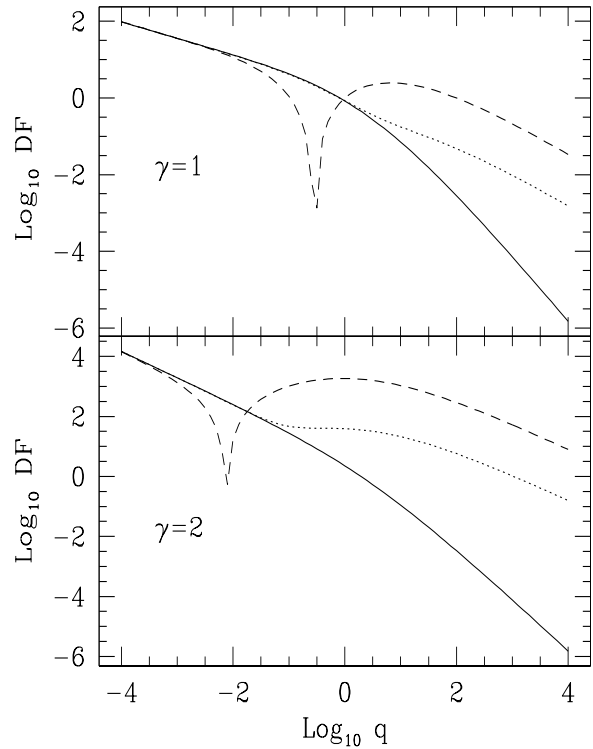}
\caption{ The phase-space DF (normalized to $\As/\vc^3\sqrt{8}\pi^2$) 
  of the stellar component of $\gamma=1$
  (top) and $\gamma=2$ (bottom) models embedded in a dark matter halo
  so that the total density profile is proportional to $r^{-2}$. Solid
  lines refer to the case of a fully isotropic stellar component,
  dotted lines to intermediate values of the (normalized) anisotropy
  radius $\sa$ ($1$ for $\gamma=1$ and $0.1$ for $\gamma=2$), and
  finally the dashed lines to a value of $\sa$ very near to the
  critical value for consistency.}
\label{fig:DF}
\end{figure}

In Fig.~\ref{fig:DF} the DFs of $\gamma=1$ and $\gamma=2$ models are
presented, in the isotropic (solid), mildly anisotropic (dotted), and
maximally anisotropic (dashed) cases. The dashed curves are
very similar to the analogous curves in Ciotti \& Lanzoni (1997,
Fig.~2), and C99 (Figs.~2 and 3), revealing the common qualitative
behavior of OM anisotropic DFs near the consistency limit,
i.e. the fact that the inconsistency manifests itself in general
at intermediate energies (see also Ciotti \& Morganti 2008 for a discussion).

\section{Jeans equations with OM anisotropy}

\subsection{Spatial velocity dispersion}

The solution of the spherical Jeans equations with general (radial or
tangential) anisotropy has been given by Binney \& Mamon (1982) and
for OM systems is given by
\begin{eqnarray}
\rhos\srad^2&=&{G\over r^2 +\ra^2}
             \int_r^{\infty}\rhos(r) \Mt(r)\left(1+{\ra^2\over r^2}\right)dr\cr
            &=&\As\vc^2 {A(s)+\sa^2I(s)\over s^2+\sa^2},
\label{eq:sigrad}
\end{eqnarray}
where $\sa=\ra/\rs$. For the present models the explicit expression of
the functions $A$ and $I$ are given in Appendix C for generic
$\gamma$, however the resulting formulae are not particularly
illuminating, as is common for this kind of models.  Nonetheless, it is
of some interest that in the $\gamma=0$ case the velocity
dispersion does not present ``unreasonable'' behaviour, even though we
know that the model is physically inconsistent.  Instead, the
asymptotic analysis of $\srad^2$ at large radii and near the center
provides helpful informations that will be used when discussing the
projected velocity dispersion profile of the models.

In the radial region $r\gg\rs$, both $A$ and $I$ can be easily
evaluted, because the stellar density profile is asymptotic to the
$r^{-4}$ profile independently of $\gamma$.  In any case $\srad^2$
tends to a constant: this is not surprising, as an elementary
integration shows that the velocity dispersion profile of power--law
densities in a total density profile $r^{-2}$ tends to a constant.  In
fact, an explicit calculation (or the expansion of eqs.~[C1]-[C3])
shows that for $s\to\infty$
\[
\srad^2\sim\vc^2{2s^2+\sa^2\over 4(s^2+\sa^2)}.
\]
Therefore, in the fully isotropic case ($\sa\to\infty$) the radial
velocity dispersion at large radii is half of the model circular
velocity. If some radial anisotropy is present, then at $r\gg\ra$ the
orbital distribution becomes fully radially anisotropic, and
accordingly the (square) intrinsic radial velocity dispersion
increases by a factor of two when compared to the isotropic case.

The situation is more delicate for $r\to 0$.  In fact, from asymptotic
expansion of the integral in eq.~(\ref{eq:sigrad}) it follows that the
central behavior of $\srad$ is coincident with that of the isotropic
case, and the product $\rhos\srad^2$ diverges for $r\to 0$
indipendently of the value of $\ra$ and $\gamma$.  In addition, for
$\ra>0$ and $\gamma>0$, the product $\rhos\srad^2$ diverges as
$\rhos$, so that $\srad^2$ converges to a finite value except for the
$\gamma=0$ models:
\[
\srad^2\sim \vc^2 
       \cases{\displaystyle{-\log s,\quad (\gamma=0)},\cr
             \displaystyle{{1\over\gamma}},\quad (0<\gamma\leq 2).}
\label{eq:sigrcen}
\]
This is relevant from the modelistic point of view, as it is well
known that self-gravitating isotropic $\gamma$ models present a
depression of their velocity dispersion near the center with
$\srad(0)=0$ (except for the $\gamma=0$ and $\gamma=2$ models, see
Bertin et al. 2002 for a general discussion of this phenomenon;
see also Binney \& Ossipkov 2001).

Before discussing the projected velocity dispersion, it can be of interest
in applications to have the analytical expression of the total kinetic
energy of the stellar component. As is well known, from the virial
theorem this quantity is independent of the specific orbital
anisotropy considered, and can be obtained without using the explicit
solution of the Jeans equations.  In fact $2K_*\equiv \int\rhos{\rm
Tr}(\sigma^2)d^3\xv= \int \langle\xv,\nabla\phit\rangle\rhos\;d^3\xv=
4\pi G \MR\Ms\As\rs^2$, where the last identity
holds for any system of finite total mass $\Ms$
in the gravitational field of the density distribution~(\ref{eq:rhot}) 
(see also Kochanek 1994), and in the present case
\[
K_*={G\Ms^2\over\rs}{(3-\gamma)\MR\over 2}.
\]
Thus, if one defines the one-dimensional stellar virial velocity
dispersion as $3\Ms\sigv^2/2=K_*$, it follows from the equation above that
\[
\sigv^2={G\Ms\over\rs}{(3-\gamma)\MR\over 3}={\vc^2\over 3},
\]
independently of the value of $\gamma$.

\subsection{Stability}

\begin{figure}
\includegraphics[scale=0.5]{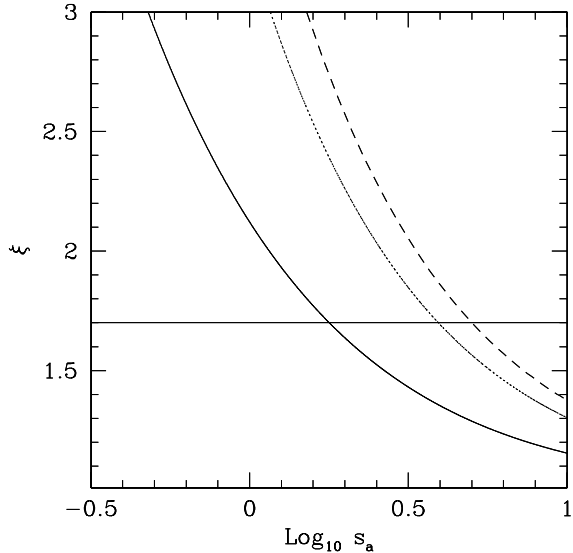}
\caption{ The stability indicator $\xi$  as a function
of the normalized anisotropy radius $\sa=\ra/\rs$, for models 
with $\gamma=1/2,1,2$ (dashed, dotted, and solid lines, respectively). 
The horizontal line marks the fiducial stability limits $\csic =1.7$. }
\label{fig:csi}
\end{figure}

{\bf Equations (29) and (32) can be used to obtain indications 
about the minimum admissible value of $\sa$ as a function of $\gamma$ 
to prevent the onset of  radial orbit instability. A complete stability 
analysis is beyond the task of this work, requiring N-body simulations or 
normal mode analysis, but some interesting conclusions can be equally derived
from the work of Fridman \& Polyachenko (1984). These authors  argued that 
a quantitative indication on the maximum amount of radial anisotropy sustainable by a specific 
density profile is given by the stability parameter $\xi\equiv 2 \Krad/\Ktan$,
where $\Krad$ and $\Ktan=K_* -\Krad$ are the radial and tangential component of
the kinetic energy tensor, respectively.  From its definition $\xi\to1$ for $\sa\to\infty$ 
(globally isotropic models), while $\xi\to\infty$ for $\sa\to 0$ (fully radially anisotropic 
models), and for one-component systems stability is associated with the empirical 
requirement that $\xi < \csic=1.7\pm 0.25$; the exact value of $\csic$ is model 
dependent (see, e.g., Merritt \& Aguilar 1985; Bertin \&
Stiavelli 1989; Saha 1991, 1992; Bertin et al. 1994; Meza \& Zamorano
1997; Nipoti, Londrillo \& Ciotti 2002). 
Here we are considering two-component systems, 
however N-body simulations have shown that the presence of a halo does not change 
very much the situation with respect to the one-component systems (e.g., 
see Stiavelli \& Sparke 1991, Nipoti et al. 2002).
Therefore, in the following discussion we assume as a fiducial maximum value for 
stability $\csic=1.7$.

The parameter $\xi$ for the present models is independent of $\MR$, and it 
cannot be expressed by using elementary functions, so that we explore 
its value numerically. In Fig.~\ref{fig:csi} we plot $\xi$
as a function of $\sa$ for $\gamma =1/2$, 1 and 2, and the asymptotic 
flattening to unity for increasing isotropy is evident. It is apparent  that the 
stability criterion requires minimum anisotropy radii appreciably larger than 
those obtained from the consistency analysis (see Sect.~3.2). 
In addition, stable stellar distributions with shallower central density profile 
require more and more isotropic velocity dispersion, 
confirming the trend already found for one and two-component 
$\gamma$-models (e.g., Carollo et al. 1995; Ciotti 1996, 1999). So, it is likely 
that the more radially anisotropic models with positive DF are prone to 
radial orbit instability.}

\subsection{Projected velocity dispersion}

The projected velocity dispersion 
associated with a general anisotropy function $\beta(r)$ 
(see eq.~[\ref{betaOM}]) is given by
\[
\Sigs(R)\sigp^2(R)= 2\int_R^{\infty}{\left [{1-\beta(r)}
{R^2\over{r^2}}\right ]}{{\rhos(r)\srad^2(r)\,r}\over{\sqrt{r^2-R^2}}}dr.
\label{eq:projsig}
\]
Unfortunately the projection integral above cannot be evaluated
analytically for generic $\gamma$ in terms of elementary functions.
However, as for the projected stellar density, interesting
informations can be obtained in two relevant radial regions, i.e.,
outside the core radius and near the center.  In practice, the
external regions are defined as the radial interval where the stellar
density profile can be approximated as a pure power--law of slope
$-4$. In this region the projection integral can be evaluated for
generic values of $\sa$ and the asymptotic result is
\[
\sigp^2(R)\sim\vc^2{(\sa^2+\eta^2)^{5/2}-\eta^3(2\sa^2+\eta^2)\over
                             4\sa^2(\sa^2+\eta^2)^{3/2}},
\label{sigRgen}
\]
where $\eta\equiv R/\rs$. In the fully isotropic case the
dimensionless ratio $\sigp(R)/\vc$ tends to 1/2, so that the projected
velocity dispersion coincides with the (constant) isotropic velocity
dispersion (as expected), while in the completely radially anisotropic
case (or for $\eta\gg\sa$) the ratio converges to $1/\sqrt{8}$.

The case of the central regions is more complicated.  In fact,
integral~(\ref{eq:projsig}) for $R\to 0$ converges when
$0\leq\gamma<1$, and diverges for $1\leq\gamma\leq2$, as can be easily
proved by using eqs.~(C5)-(C6).  In the divergent case both
the projection integral and the projected surface density $\Sigs$ are
asymptotically dominated by their integrands for $r\sim0$ (as is
intuitive, the cusp dominates and the contribution of foreground stars
and background stars is negligible), and $\sigp(R)$ can be properly
defined as the limit value of their ratio for $R\to 0$.  A simple
calculation shows that
\[
\sigp(0)=\srad(0)={\vc\over\sqrt{\gamma}},\quad 1\leq\gamma\leq 2.
\label{sig0div}
\]
In other words, for the models with the centrally divergent surface
(stellar) density, {\it the projected central velocity dispersion
coincides with the central radial component of the isotropic velocity
dispersion}. Thus it does not depend
on the model anisotropy radius, at variance with 
the projected velocity dispersion in the external galactic regions.

In the convergent case $0\leq\gamma<1$ the central value of the
projection integral depends on the whole profile of the integrand,
therefore the projected central velocity dispersion depends also on
$\sa$. For generic $\gamma$ and $\sa$, the quantity $\sigp(0)$ is
expressible in terms of hypergeometric ${}_2F_1$ functions.  However a
simple form of the projection integral evaluated at $R=0$, useful in
numerical integrations, can be obtained by inverting the order of
integration in eq.~(\ref{eq:projsig}):
\begin{eqnarray}
\label{sig0conv}
&&\Sigs(0)\sigp^2(0)= 2\int_0^{\infty}\rhos(r)\srad^2(r)\,dr=
{8\pi G\MR\As^2\rs^3\over\sa}\times\cr
&&
\int_0^{\infty}{s^{1-\gamma}\over(1+s)^{4-\gamma}}
\left(1+{\sa^2\over s^2}\right)\arctan{s\over\sa}ds.
\end{eqnarray}
In the fully isotropic case ($\sa\to\infty$) the projection integral can be
evaluated analytically and the result is
\[
\sigp(0)=\vc,\quad 0\leq\gamma <1,
\label{sig0con}
\]
which is independent of $\gamma$. This is not surprising, as it is
easy to show, by inverting order of integration, that for {\it any}
density profile with finite central projected density, isotropic
orbital distribution, in the total gravitational field produced by
density distribution (6), identity (38) holds.

Equations~(\ref{sigRgen}), (\ref{sig0div}), and (\ref{sig0con}) open
the interesting possibility to consider the ratio of the outer to the
central projected velocity dispersion, a quantity that can be
expressed in a very simple way.  In particular, the ratio depends on
the shape of the stellar density slope $\gamma$, on the outer
observational point $R$, and finally on $\sa$. Thus, at least in
principle, for galaxies well described by a $\gamma$-model immersed in
a total density profile $\propto r^{-2}$, it could be possible to
determine $\sa$ from observations, assuming OM anisotropy.  In Fig.~4
the ratio is plotted as a function of the anisotropy radius for the
representative values of the density slope $2,1$, and $1/2$ (in this
latter case the WSC limit for consistency is $\sa\gsim0.45$).  All the
expected trends are apparent, iin particular the decrease of the ratio
$\sigp(R)/\sigp(0)$ for decreasing $\sa$. This is due, for
$1\leq\gamma\leq 2$, to the decrease of $\sigp(R)$ in the external
regions of radially anisotropic models. In the $\gamma =1/2$ case,
reported as an example of models with central slope in the range
$0\leq\gamma <1$, the larger decrease is due to the additional effect
of the increase of $\sigp(0)$ for decreasing $\sa$. Overall, the
kinematical ratio $\sigp(R)/\sigp(0)$ for mildly-strongly anisotropic
models (i.e., $\sa\lsim 1$) is $\sim 15\% -20\%$ lower than in the
corresponding isotropic cases.

\subsection{Some additional considerations}

For sake of completeness, we summarize some additional results 
on velocity dispersion. For example, the central velocity dispersion 
of $\gamma$ models in the presence of a black hole can be found 
in the literature.  Here we just recall that the velocity dispersion 
diverges for $r\to0$ as $r^{-1/2}$ (e.g., see C96, Baes \&
Dejonghe 2004, Baes et al. 2005; for the case of oblate $\gamma$
models, or two-component oblate power-law models with central black
hole see also Riciputi et al. 2005, Ciotti \& Bertin 2005). It follows
that in the present context a central black hole would produce an
identical kinematical signature, as sufficiently near the center the
total mass is fully dominated by the black hole.

As a second case, we consider the spatial and projected velocity
dispersion of a two-component galaxy model made by the superposition
of a stellar component described by a $\gamma$ model (where for
simplicity we restrict to the interval $1\leq\gamma\leq2$), and a dark
halo component described by a Jaffe model ($\gamma=2$).  The interest
of these models is due to the fact that in the inner regions they
behave as the models subject of this work, but in the external regions
a Keplerian decline is present, due to the halo finite total mass.
From eq.~(\ref{eq:rhos}), the Jaffe profile of total mass $M_{\rm
  h}=\MR\Ms$ and scale length $\rh=\beta\rs$ can be written as
\[
\rhoh(r)={\As\MR\beta\over (3-\gamma)\,s^{2}(\beta+s)^{2}}.
\label{eq:rhoj}
\]
The Jeans equations for this class of models cannot be solved
explicitly in terms of elementary functions for generic $\gamma$, even
though special explicit cases can be easily found (e.g., see Ciotti et
al. 1996).  For this reason we just evaluate the asymptotic leading
term of the velocity dispersion for $s\to\infty$, obtaining
\[
\srad^2\sim{G\Ms(1+\MR)\over\rs}{5s^2+3\sa^2\over 15s(s^2+\sa^2)},
\]
and its projection through eq.~(\ref{eq:projsig}), giving
\begin{eqnarray}
\sigp^2(R)&\sim&{G\Ms\over\rs}{4(1+\MR)\over15\pi\eta}
\left[2+{\eta^4\over\sa^2(\sa^2+\eta^2)}\right.\nonumber\\
&-&\left.{\eta^4(2\sa^2+\eta^2)
\sinh^{-1}(\sa/\eta)\over\sa^3(\sa^2+\eta^2)^{3/2}}\right],
\end{eqnarray}
where $\eta\equiv R/\rs$. The expression in square parentheses 
converges to $2$ in the isotropic case.

The behavior of the velocity dispersion in the central regions requires
some additional discussion.  In fact, for $1\leq\gamma\leq2$ the
central projected velocity dispersion will depend on the central,
isotropic spatial velocity dispersion only (for the reasons described
in Section 4.2, and excluding the purely radial case). In addition,
the self-contribution to the central velocity dispersion of the
$\gamma$ model is zero (except for the $\gamma =2$ case, e.g., see
Bertin et al. 2002).  Summarizing, the projected central velocity
dispersion of these models is coincident with the isotropic spatial
velocity dispersion of $\gamma$ models in the presence of the Jaffe
halo only (with the exception of the $\gamma=2$ case).  Therefore,
\[
\sigp^2(0)=\srad^2(0)={G\Ms\over\rs} 
       \cases{\displaystyle{{\MR\over\beta\gamma}},\quad\quad(1\leq\gamma<2),\cr
              \displaystyle{{\beta+\MR\over2\beta},\quad (\gamma=2).}}
\label{eq:sigrcenJaf}
\]
\subsection{Velocity profile}
Not surprisingly, the velocity profile VP cannot be expressed in terms
of elementary functions, however acceptably simple formulae can be
obtained at large radii and in the central galactic regions.  The
starting point is to consider the normalization of eq.~(\ref{eq:vp}),
given by
\begin{equation}
\label{eq:vp1}
\Sigs\mbox{VP}=-{\sqrt{2}\over\pi}{\As\rs\over\vc}
                \int^{\infty}_\eta {g(s,\eta)sG(\qm)ds\over\sqrt{s^2-\eta^2}},
\end{equation}
where $\qm\equiv\Qm/\vc^2$, and
\begin{equation}
G(\qm)=\int_{\qm}^{\infty}{d\tilde\varrho\over d\Psi}
     {d\Psi\over\sqrt{\Psi-\qm}}.
\end{equation}
At the very center of the stellar system, where the stellar density
profile is proportional to $s^{-\gamma}$, it is not difficult to show
that for $1\leq\gamma\leq2$ 
\begin{equation}
\mbox{VP}\sim\sqrt{\gamma\over2\pi\vc^2}\,e^{-\gamma\vpar^2/2\vc^2}
\end{equation}
independently of the (positive) value of the anisotropy
radius. Instead, for $0\leq\gamma <1$, as for the projected velocity
dispersion a numerical integration is required.

In the external galactic regions, where the stellar density profile
can be approximated as $s^{-4}$ power--law, the VP can be written as a
quite simple integral, depending on the dimensionless ratios
$\vpar/\vc$ and $\eta/\sa$, that can be evaluated numerically.  Here
we just report the formula for the isotropic case, where
\begin{equation}
\mbox{VP}\sim\sqrt{2\over\pi\vc^2}\,e^{-2\vpar^2/\vc^2}.
\end{equation}
In both the reported formulae, the Gaussian signature is apparent.
\begin{figure}
\includegraphics[scale=0.5]{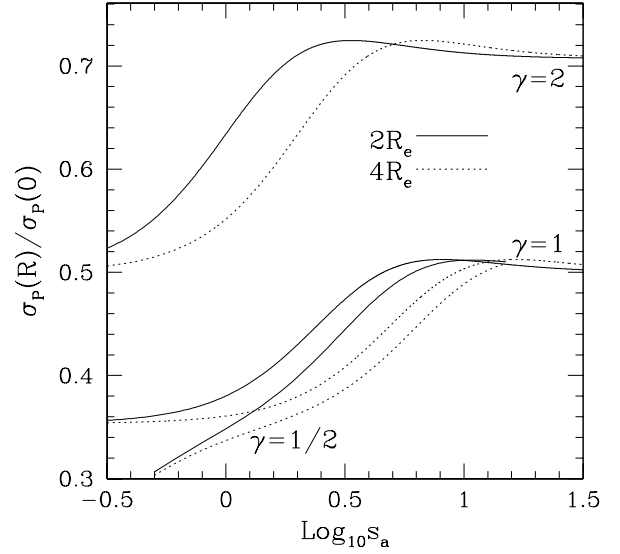}
\caption{ The ratio $\sigp(R)/\sigp(0)$ as a function
of the normalized anisotropy radius $\sa=\ra/\rs$, for models 
with $\gamma=1/2,1,$ and $2$.
Solid lines refer to $R=2\reff$, dotted lines to $R=4\reff$.}
\label{fig:jump}
\end{figure}
\section{Conclusions}

In this paper a family of spherical, two-component galaxy models with
a stellar density profile described by a $\gamma$ model and a total
(stars plus dark matter) density profile $\propto r^{-2}$ at all radii
has been investigated, under the assumption that the internal dynamics
of the stellar component is described by Osipkov-Merritt anisotropy.
The models are fully determined when the inner density slope $\gamma$
and the anisotropy radius $\ra$ of the stellar component are assigned,
together with a density scale for the total density profile.  The dark
matter halo remains defined as the difference between the total and
stellar density profiles. The main results can be summarized as
follows.

\begin{itemize}

\item After having provided the most common structural quantities of
  the models that are of interest for observations, limitations on the
  {\it total} density scale as a function of $\gamma$ are analytically
  determined by requiring the positivity and monotonicity of the dark
  matter halo distribution.  In particular, the request of positivity
  limits the range of acceptable stellar density slopes to
  $0\leq\gamma\leq 2$.  Models corresponding to the minimum total
  density scale (for given $\gamma$) are called minimum halo models.
  The central density profile of the dark matter halo diverges as
  $r^{-2}$ in general, but in the minimum halo $\gamma=2$ model (in
  which the positivity and monotonicity limits coincide), the central
  dark matter profile is $\propto r^{-1}$.

\item The minimum value of anisotropy radius corresponding to a
  dynamically consistent stellar component has been derived
  analytically as a function of $\gamma$ by using the necessary and
  sufficient conditions of CP92.  As expected, an increase of $\gamma$
  results in a decrease of the minimum value of the anisotropy radius
  $\ra$ required by consistency.  It is also proved that models with
  $\gamma>4/17$ are certainly consistent for sufficiently isotropic
  velocity dispersion, while for centrally shallower stellar density
  profiles the sufficient condition for consistency is never
  satisfied.  The necessary and sufficient condition for the halo
  consistency are also analytically obtained, and the minimum halo
  models corresponding to (isotropic) dark matter halos are derived.
  In the case $\gamma=2$, the minimum halo coincides with the minimum
  halo obtained from the positivity and monotonicity conditions.

\item The phase--space DF of the stellar component for $\gamma=0,1,2$
  is analytically recovered in terms of Polylogarithms and
  exponentials.  It is found that the $\gamma=0$ model is inconsistent
  no matter how much anisotropy is considered. Instead, the isotropic
  $\gamma=1,2$ models have a positive DF, and the true critical
  anisotropy radius for consistency can be determined directly from
  their DF.  A comparison with the analogous study of one-component
  $\gamma$ models shows that the presence of the halo sligthly reduces
  the maximum amount of sustainable radial anisotropy. The obtained
  values of the anisotropy radius are independent of the total density
  scale $\MR$.

\item The Jeans equations for the stellar component are solved
  explicitely for generic values of $\gamma$ and $\ra$ in terms of
  elementary functions.  The asymptotic expansions of $\srad$ for
  $r\to0$ and $r\to\infty$ are obtained, and it is shown that $\srad$
  tends to finite (non-zero) values (except for the divergent central
  velocity dispersion of $\gamma=0$ model) which are simply related to
  the model circular velocity.  The projected velocity dispersion
  $\sigp(0)$ cannot be calculated analytically, in general. However,
  by asymptotic expansion of the projection integral, exact values at
  large radii and at the center are obtained.  In particular, it is
  shown that for $\gamma\geq 1$, and independently of the value of the
  anisotropy radius, $\sigp(0)$ coincides with the central velocity
  dispersion $\srad(0)$ in the isotropic case.  Instead, for
  $0\leq\gamma<1$, $\sigp(0)$ depends also on $\sa$.  In the
  anisotropic case $\sigp(0)$ cannot be obtained analytically;
  however, a very simple form of the projection integral suitable for
  numerical integrations is given. In the isotropic case the integral
  can be evaluated analytically, and $\sigp(0)$, as expected,
  coincides with the model circular velocity.

\item Finally, we have shown that the Velocity Profile of the models
  can be obtained in a very simple form (Gaussian) near the center
  (independently of the value of the anisotropy radius and for
  $1\leq\gamma\leq 2$), and at large radii (in the isotropic case).

\end{itemize}

We conclude by noting that these models, albeit highly idealized, seem
to suggest two interesting remarks of observational character.  The
first is that in real galaxies with a total $r^{-2}$ density profile
and sufficiently peaked stellar density (i.e. $\gamma\geq1$), measures
of central velocity dispersion should not be strongly affected by
radial anisotropy.  Second, for given central stellar density slope
$\gamma$, measures of the projected velocity dispersion at the center
and in the external regions are able, at least in principle, to
determine the value of the anisotropy radius under the assumption of
Osipkov-Merritt anisotropy.

\section*{Acknowledgments}
L.C. wishes to thank Carlo Nipoti and Leon Koopmans for useful comments.
{\bf We also thank an anonymous referee for useful comments.}

\appendix

\section{Necessary and sufficient conditions for model consistency}

The condition for the {\it positivity} of the halo density profile $\rhoh$
can be easily established from eq.~(\ref{eq:rhoh}). In fact, for
$0\leq\gamma\leq 2$, $\MR$ must be greater or equal to the maximum of
the radial function inside the parentheses, and simple algebra shows
that the maximum is attained for
\[
s_m={2-\gamma\over 2};
\]
in particular, for the Jaffe model the critical point is reached at
the center.  The {\it monotonicity} condition is obtained requiring
that the radial derivative of $\rhoh$ is nowhere positive, and this
happens if and only if
\[
\MR\geq {s^{2-\gamma}(\gamma+4s)\over 2(1+s)^{5-\gamma}}\quad\forall s\geq 0.
\label{eq:mona}
\]
Thus, 
$\MR$ must be greater than or equal to the maximum of the
radial function above, that is reached at
\[
s_m={12-7\gamma+\sqrt{(4-\gamma)(36-17\gamma)}\over 16},
\]
and again for $\gamma=2$ the maximum is reached at the center. Finally, 
the application of the WSC to an isotropic halo in order to have phase-space 
{\it consistency} is given by eq.~(\ref{eq:WSC}) with $\ra\to\infty$, 
so that $\varrho=\rhoh$. The condition becomes
\[
\MR\geq {s^{2-\gamma}[16s^2+(9\gamma-4)s+\gamma^2]\over 4(1+s)^{6-\gamma}}
\quad\forall s\geq 0.
\label{eq:wscha}
\]
The study of the maximum of the r.h.s.  of equation above leads to
a cubic equation.  We do not report here the solution
$s_m(\gamma)$ corresponding to the maximum, as it can be easily
obtained, but for reference we just report three special values
$s_m(0)\simeq 2.2049$ $s_m(1)\simeq 1.2079$ and $s_m(2)=0$,
corresponding to $\MR_m(0)\simeq 0.07735$, $\MR_m(1)\simeq 0.17487$,
and $\MR_m(2)=1$, respectively.  

The application of the WSC to the stellar component reduces instead to 
the study of 
\[
16s^2+(9\gamma-4)s+\gamma^2+
s^2{4s^2+(5\gamma-12)s+(\gamma-2)^2\over\sa^2}\geq 0, 
\label{eq:condm}
\]
so that the function $F$ in eq.~(\ref{eq:cond}) is a quadratic
polynomial, and the determination of the sets $\condp$ and $\condm$ is
straigthforward.  In particular, $\condm$ is not empty for
$0\leq\gamma\leq 4/17$, and the function $G$ is positive there.

\section{Phase-space DF for the stellar component}

Here we give the explicit evaluation of the DF for the three integer
values $\gamma=0,1,2$. In fact, in these cases one can change the
variable integration by defining $t=\sqrt{\Psi-q}$, so that
$e^{\Psi}=e^{t^2+q}$ and the integration interval is mapped into
$(0,\infty)$. Expansion in simple fractions, factorization of $e^q$
outside the integrals and repeated differentiation with respect to
$e^{-q}$ under the sign of the integral finally shows that for
$\gamma=0$
\begin{eqnarray}
&&U(q)={\sqrt{\pi}\over 6}\times\cr
    &&\left[\Lin_{-9/2}(y)-6\Lin_{-7/2}(y)+
            11\Lin_{-5/2}(y)-6\Lin_{-3/2}(y)\right],
\end{eqnarray}
\[
V(q)={\sqrt{\pi}\over 6}\times\left[\Lin_{-9/2}(y)-\Lin_{-5/2}(y)\right],
\]
where $y\equiv -e^{-q}$, $\Lin_s(z)=z\Phi(z,s,1)$ is the so-called
Polylogarithm function, $\Phi(z,s,a)$ is the Lerch function and
$d\Lin_s(z)/dz=\Lin_{s-1}(z)/z$ (e.g., Erd\'elyi et al. 1953).  A
similar treatement for the $\gamma=1$ case gives
\begin{eqnarray}
U(q)&=&{\sqrt{\pi}\over 2}\times\cr
    &&\left[e^{-q}+\Lin_{-7/2}(y)-5\Lin_{-5/2}(y)+6\Lin_{-3/2}(y)\right],
\end{eqnarray}
\[
V(q)={\sqrt{\pi}\over 2}\times\left[\Lin_{-7/2}(y)-\Lin_{-5/2}(y)\right],
\]
and finally for $\gamma=2$
\[
U(q)=\sqrt{\pi}\times
\left[2^{3/2}e^{-2q}-2e^{-q}+\Lin_{-5/2}(y)-3\Lin_{-3/2}(y)\right],
\]
\[
V(q)=\sqrt{\pi}\times\left[\Lin_{-5/2}(y)-\Lin_{-3/2}(y)\right].
\]

\section{Velocity dispersions}

The isotropic function $I$ in eq.~(\ref{eq:sigrad}) 
for $\gamma\neq 0,1,2$ is given by
\begin{eqnarray}
I&=&{6s^3+6(3-\gamma)s^2+3(3-\gamma)(2-\gamma)s+
   (3-\gamma)(2-\gamma)(1-\gamma)\over 
   s^{\gamma}(1+s)^{3-\gamma}(3-\gamma)(2-\gamma)(1-\gamma)\gamma}\cr
&& -{6\over (3-\gamma)(2-\gamma)(1-\gamma)\gamma},
\end{eqnarray}
while
\begin{eqnarray}
I=\cases{\displaystyle{\log {1+s\over s}-{11+15s+6s^2\over 6(1+s)^3}},
          \quad\gamma=0\cr
         \displaystyle{{1\over s}+{5+4s\over 2(1+s)^2} -3\log {1+s\over s}},
          \quad\gamma=1\cr
         \displaystyle{3\log {1+s\over s}+{1-3s-6s^2\over 2s^2(1+s)}},
          \quad\gamma=2.}
\end{eqnarray}
The function $A$ is given by
\begin{eqnarray} 
A=\cases{\displaystyle{{1\over (3-\gamma)(2-\gamma)}- 
                       {(3-\gamma+s)s^{2-\gamma}\over 
                        (3-\gamma)(2-\gamma)(1+s)^{3-\gamma}},
                      }\cr
         \displaystyle{\log{1+s\over s}-{1\over 1+s},\quad \gamma =2.}
        }
\end{eqnarray}
For $s\to\infty$ and $0\leq\gamma\leq 2$
\[
I\sim {1\over 4s^4},\quad A\sim {1\over 2s^2},
\]
while for $s\to 0$
\begin{eqnarray}
I\sim\cases{\displaystyle{-\log s}, \quad\gamma=0\cr
         \displaystyle{1\over \gamma s^{\gamma}},
          \quad 0<\gamma\leq 2,}
\label{eq:iszero}
\end{eqnarray}
and
\begin{eqnarray}
A\sim\cases{\displaystyle{1\over (3-\gamma)(2-\gamma)}, \quad 0\leq\gamma <2\cr
         \displaystyle{-\log s}, \quad\gamma =2.}
\label{eq:aniszero}
\end{eqnarray}


\end{document}